\documentclass[12pt]{article}

\usepackage{amsmath}
\usepackage{amssymb}
\usepackage[pdftex]{graphicx}
\usepackage{amscd}
\usepackage{pictexwd,dcpic}
\usepackage{slashed}
\usepackage{mathtools}
\usepackage{braket}

\numberwithin{equation}{section}

\begin{document}

\begin{flushright}
UTTG-13-16
\end{flushright}

\begin{center}

\LARGE{Topological Sigma Models On Supermanifolds}

\large{Bei Jia}

\footnotesize{Theory Group, University of Texas at Austin, Austin, TX 78712, USA}

\footnotesize{beijia@physics.utexas.edu}

\end{center}

\date{}


\begin{abstract}
This paper concerns constructing topological sigma models governing maps from semirigid super Riemann surfaces to general target supermanifolds. We define both the A model and B model in this general setup by defining suitable BRST operators and physical observables. Using supersymmetric localization, we express correlation functions in these theories as integrals over suitable supermanifolds. In the case of the A model, we obtain an integral over the supermoduli space of ``superinstantons''. The language of supergeometry is used extensively throughout this paper.
\end{abstract}

\newpage

\tableofcontents

\newpage

\section{Introduction} \label{intro}

Topological field theories are special quantum field theories whose correlation functions are topological and can be solved exactly. They have been discussed numerous times over the years and have a rather long list of applications. Their main properties can even be turned into rigorous mathematical definitions, which is a very rare phenomenon in physics.

The A model and B model topological sigma models, introduced by Witten decades ago \cite{witten-tft}, provide important concrete realizations of abstract topological field theories. As quantum field theories, they govern maps from Riemann surfaces to target manifolds. These theories have their physical spectra defined by cohomologies of suitable nilpotent BRST operators. Correlation functions can be, in principle, calculated exactly at the full quantum level. In particular, correlation functions in the ordinary A model defines quantum cohomology rings on target manifolds.

The aim of this paper is the establish basic constructions and analysis of a generalization of ordinary topological sigma models. We will define our version of topological sigma models as governing maps from semirigid super Riemann surfaces \cite{distler-nelson, distler-nelson-2, mine-ts} to target supermanifolds. In other words, as quantum field theories they depend on maps between supermanifolds. One major effort is to use techniques from supergeometry extensively. This is inspired by the recent development in superstring perturbation theory and topological string theory \cite{witten-spt, mine-ts}. In our approach, things are naturally defined using properties of supermanifolds. Ordinary topological sigma models can be obtained as special cases in which target spaces are taken to be ordinary even manifolds.

Topological sigma models with supermanifolds as target spaces have been studied before, see for example \cite{sethi, schwarz, witten-twistor}. In particular, in \cite{witten-twistor} it was shown that B model with target Calabi-Yau suparmanifold $\mathbb{CP}^{3|4}$ has deep relation to amplitudes in $\mathcal{N}=4$ super Yang-Mills theory. However, those studies are rather incomplete and not systematic. One basic difference of our approach with e.g. \cite{sethi, schwarz} is that they are in some sense ``half-way'' in between ordinary topological sigma models and topological sigma models defined in this paper. For example, they primarily consider maps from ordinary Riemann surfaces to supermanifolds.

The structure of this paper is the following. In section \ref{basics} we laid down some basic definitions and properties of our topological sigma models in general. We review the concept of semirigid super Riemann surfaces, and construct topological sigma model worldsheets based on them. Then in section \ref{A} and \ref{B} we study the A model and B model on supermanifolds systematically. We define the appropriate BRST operators, construct worldsheets, analyze anomalies and define physical observables. In the A model, correlation functions between physical operators get quantum corrections, as these correlation functions are localized to integrals over some supermoduli spaces. In contrast, correlation functions in the B model are purely classical, localized to integrals over the target supermanifolds. Finally, in section \ref{discussion} we discuss many open questions naturally raised by our construction.

There are also some relevant appendices at the end of this paper. In appendix \ref{maps}, we review some basics of supermanifolds in general and maps between them. In appendix \ref{forms}, some basic facts about pseudoforms on supermanifolds are given. In appendix \ref{susy-transf} we explicitly work out actions of supercharges and supersymmetry transformations.

\section{The Basics} \label{basics}

\subsection{Semirigid Super Riemann Surface}

Complex supermanifolds of dimension $1|m$ are of special interests to string theory and related topics, as they can be identified as the worldsheet of various objects. When $m = 1$, one can further impose $\mathcal{N} = 1$ superconformal structure which provides worldsheets for RNS superstrings. In \cite{distler-nelson, distler-nelson-2, mine-ts}, it was argued that semirigid super Riemann surfaces should naturally serve as worldsheets for topological string theory, which naturally requires $m = 2$.

In principle, topological string theory contains topological field theory as ``matter''. Therefore, as a priori semirigid super Riemann surfaces should provide natural background for studying two-dimensional topological field theories. In this section, we will review some basic information about semirigid super Riemann surfaces.

A semirigid super Riemann surface can be constructed by ``twisting'' a $\mathcal{N} = 2$ super Riemann surface, in a rather analogous way like the usual topological twistings used to define topological field theoreis \cite{distler-nelson, distler-nelson-2, mine-ts}. Here we will not go through that route; instead we will define these surfaces by the natural data that resides on them.

A semirigid super Riemann surface $\mathbf{S}_-$ is a complex supermanifold of dimension $1|2$, with the following data:
\begin{itemize}
\item Two odd subbundles, $\mathcal{D}_-$ and $\mathcal{D}_+$, of the holomorphic tangent bundle $T\mathbf{S}_-$ of $\mathbf{S}_-$;
\item Sections of $\mathcal{D}_- \otimes \mathcal{D}_+$ are everywhere linearly independent of sections of $\mathcal{D}_-$ and $\mathcal{D}_+$;
\item $\mathcal{D}_-$ and $\mathcal{D}_+$ are integrable;
\item We trivialize $\mathcal{D}_-$, then assign spin 0 to $\theta^-$ and spin 1 to $\theta^+$.
\end{itemize} 
Note the first three conditions define general $\mathcal{N}=2$ super Riemann srufaces. Let $(z|\theta^-,\theta^+)$ be local coordinates on $\mathbf{S}_-$. Integrability here means that, for any given sections $D_{\mp}$ of $\mathcal{D}_{\mp}$, there exists functions $f_-(z|\theta^-,\theta^+)$ and $f_+(z|\theta^-,\theta^+)$ such that 
\begin{equation}
D_{\mp}^2 = f_{\mp}(z|\theta^-,\theta^+) D_{\mp}.
\end{equation}

It can be shown that there exits a set of local coordinates $(z|\theta^-,\theta^+)$ such that sections $D_{\mp}$ of $\mathcal{D}_{\mp}$ can be expressed as 
\begin{equation}
\begin{split}
D_- &= \frac{\partial}{\partial\theta^-} + \theta^+ \frac{\partial}{\partial z} \\
D_+ &= \frac{\partial}{\partial\theta^+} + \theta^- \frac{\partial}{\partial z} 
\end{split}
\end{equation}
One might recognize these expressions from the usual construction of supersymmetry in two dimensions. They satisfy the usual relations:
\begin{equation}
D_-^2 = D_+^2 = 0, \ \ \ \ \{D_-, D_+\} = 2 \partial_z
\end{equation}

Clearly, we could have chosen to trivialize $\mathcal{D}_+$, with $\theta^-$ having spin 1 and $\theta^+$ having spin 0. This way we obtain an isomorphic surface which we call $\mathbf{S}_+$. At this point, this choice seems rather arbitrary and inconsequential. However, later we will see the effect of making different choices.

Let's concentrate on $\mathbf{S}_-$ for the moment. Then as a semirigid super Riemann surface,  $\mathbf{S}_-$ can constructed by glueing together open subsets of $\mathbb{C}^{1|2}$ by the following coordinate transformations \cite{distler-nelson, mine-ts}:
\begin{equation} \label{semirigid-transf}
\begin{split}
z' & = f(z) + \theta^-  \rho(z), \\
{\theta^-}^{\prime} & = \theta^-,   \\
{\theta^+}^{\prime} & = \rho(z) + \theta^+ \partial_z f(z) - \theta^- \theta^+ \partial_z \rho(z),
\end{split}
\end{equation}
where $f(z)$ is an even function of $z$ and $\rho(z)$ is an odd function of $z$. We call these transformations semirigid coordinate transformations. Note that $\theta^-$ transforms trivially. Therefore, our assignment of spin 0 to $\theta^-$ is valid.

These semirigid coordinate transformations are generated by certain vector fields on $\mathbf{S}_-$, called semirigid vector fields \cite{mine-ts}. A particular basis for them can be expressed as:
\begin{equation} \label{semirigid-vector}
\begin{split}
\mathcal{T}_- & =  g(z)  \partial_z + \partial_z g(z) \theta^+\partial_{\theta^+}, \\
\mathcal{J} & = k(z)  (\theta^-\partial_{\theta^-} - \theta^+\partial_{\theta^+}), \\
\mathcal{G}^- & =\alpha^- (\partial_{\theta^-} - \theta^+ \partial_z), \\
\mathcal{G}^+ & = \left( \alpha^+(z) -  \theta^-\theta^+ \partial_z \alpha^+(z) \right) \partial_{\theta^+} -  \alpha^+(z) \theta^- \partial_z,
\end{split}
\end{equation}
where $g(z)$ and $k(z)$ are even functions of $z$ such that $g - k$ is a constant, $\alpha^-$ is an odd constant, and $\alpha^+(z)$ is an odd function of $z$. Note that $\mathcal{G}^- = \alpha^- \mathcal{Q}_-$, where 
\begin{equation}
\mathcal{Q}_- = \frac{\partial}{\partial\theta^-} - \theta^+ \frac{\partial}{\partial z},
\end{equation}
is what's usually called a supersymmetry charge in the literature.

Analogously, there is a set of corresponding vector fields on $\mathbf{S}_+$ with similar expressions. In particular, there is a nilpotent supersymmetry charge
\begin{equation}
\mathcal{Q}_+ = \frac{\partial}{\partial\theta^+} - \theta^- \frac{\partial}{\partial z},
\end{equation}
on $\mathbf{S}_+$. These nilpotent charges $\mathcal{Q}_{\mp}$ are essential for defining cohomological topological field theories.

Complex functions on $\mathbf{S}_-$ are of special interests to physics. They are usually called $(0,2)$ superfields in the physics literature. Let $\mathbf{F}: \mathbf{S}_- \rightarrow \mathbb{C}$ be such a function. If $D_- \mathbf{F} = 0$, then $\mathbf{F}$ is called a (0,2) chiral superfield. Analogously, if $D_+ \mathbf{F} = 0$, then $\mathbf{F}$ is called a (0,2) antichiral superfield. These are useful for constructing ``half-twisted'' topological sigma models with ordinary bosonic target spaces.

Finally, we would like to stress a special property of semirigid super Riemann surfaces: the Berezinian line bundle of a semirigid super Riemann surface is trivial \cite{mine-ts}. This fact leads to our later construction of Lagrangians of topological sigma models. More precisely, the trivialness of the Berezinian line bundle here implies that Lagrangians of any quantum filed theory defined on such surfaces will have their Lagrangian being global functions.

\subsection{Topological Sigma Models}

In this section we will generalize the usual topological sigma models, namely the A model and B model, to have supermanifolds as target spaces. First of all, we need to define a natural playground for them. This can be most conveniently achieved by using the construction introduced in \cite{witten-spt} for defining superstring worldsheets.

Recall that two dimensional field theories have two independent degrees of freedom: right-moving and anti-moving. In terms of complex coordinates, these correspond to holomoprhic and anti-holomorphic quantities, respectively. To characterize them, we include two copies of semirigid super Riemann surfaces, $\mathbf{S}_R$ and $\mathbf{S}_L$, and consider the diagonal
\begin{equation}
\mathbf{\Sigma} \hookrightarrow \mathbf{S}_L \times \mathbf{S}_R
\end{equation}
In this way, holomorphic quantities on $\mathbf{S}_R$ define holomorphic quantities on $\mathbf{\Sigma}$, while holomorphic quantities on $\mathbf{S}_L$ define anti-holomorphic quantities on $\mathbf{\Sigma}$. In particular, on $\mathbf{\Sigma}$ we have holomorphic odd bundles $\mathcal{D}_{\mp}$ and anti-holomorphic odd bundles $\overline{\mathcal{D}}_{\mp}$.

There are two basic types of such backgrounds. If we have $\mathbf{S}_R = \mathbf{S}_+$ and $\mathbf{S}_L = \mathbf{S}_+$, then the resulting worldsheet is called the A-type, denoted as $\mathbf{\Sigma}_A$. In contrast, if both $\mathbf{S}_R$ and $\mathbf{S}_L$ are of the type $\mathbf{S}_-$, then the resulting worldsheet is called the B-type, denoted as $\mathbf{\Sigma}_B$. As these names suggest, $\mathbf{\Sigma}_A$ and $\mathbf{\Sigma}_B$ are natural backgrounds for the A model and B model, respectively. 

In most of this section, we will focus on some general properties of both $\mathbf{\Sigma}_A$ and $\mathbf{\Sigma}_B$, so we will not distinguish between them and use the notation $\mathbf{\Sigma}$ to denote both of them. However, it is important to keep in mind one subtlety: on $\mathbf{\Sigma}_A$, $\mathcal{D}_{\mp}$ are complex conjugate to $\overline{\mathcal{D}}_{\pm}$, while on $\mathbf{\Sigma}_B$, $\mathcal{D}_{\mp}$ are complex conjugate to $\overline{\mathcal{D}}_{\mp}$.
 
As in any quantum field theory, the first step towards constructing a theory is to define a proper field content. Intuitively, we should consider complex functions on $\mathbf{\Sigma}$. In particular, to construct topological sigma models, we should consider chiral superfields, namely complex functions $\Phi: \mathbf{\Sigma} \rightarrow \mathbb{C}$ such that
\begin{equation}
D_- \Phi = \overline{D}_- \Phi = 0.
\end{equation}
Analogously, we call a function $\overline{\Phi}: \mathbf{\Sigma} \rightarrow \mathbb{C}$ an antichiral superfield if 
\begin{equation}
D_+ \overline{\Phi} = \overline{D}_+ \overline{\Phi} = 0.
\end{equation}

Locally, at any point $(z,\bar{z}|\theta^{\mp},\bar{\theta}^{\mp})$ on $\mathbf{\Sigma}$, we can expand these functions as usual:
\begin{equation}
\begin{split}
\Phi  =& \ \phi + \theta^+ \chi + \bar{\theta}^+ \lambda + \theta^+ \bar{\theta}^+ F - \theta^-\theta^+ \partial_z \phi - \bar{\theta}^- \bar{\theta}^+ \partial_{\bar{z}}\phi \\
& - \theta^+ \bar{\theta}^- \bar{\theta}^+ \partial_{\bar{z}} \chi - \theta^- \theta^+ \bar{\theta}^+ \partial_z\lambda + \theta^- \theta^+ \bar{\theta}^- \bar{\theta}^+ \partial_z \partial_{\bar{z}} \phi, \\
\overline{\Phi} =& \ \bar{\phi} + \theta^- \bar{\chi} + \bar{\theta}^- \bar{\lambda} +  \theta^- \bar{\theta}^- \bar{F}  + \theta^-\theta^+ \partial_{z} \bar{\phi} + \bar{\theta}^- \bar{\theta}^+ \partial_{\bar{z}}\bar{\phi} \\
&+ \theta^- \bar{\theta}^- \bar{\theta}^+ \partial_{\bar{z}} \bar{\chi} + \theta^- \theta^+ \bar{\theta}^- \partial_{z}\bar{\lambda} + \theta^- \theta^+ \bar{\theta}^- \bar{\theta}^+ \partial_z \partial_{\bar{z}} \bar{\phi}.
\end{split}
\end{equation}
These component fields have various geometric meanings, depending whether we are dealing with $\mathbf{\Sigma}_A$ or $\mathbf{\Sigma}_B$. The details will be discussed later separately for A model and B model. Here we only remark that the lowest components, namely $(\phi, \bar{\phi})$, are usually interpreted as local coordinates parametrizing the target space here, which is simply $\mathbb{C}$.

Now suppose we have many such chiral superfields, indexed as $\Phi^i$, and antichiral superfields, indexed as $\overline{\Phi}^{\bar{\imath}}$. In addition, suppose we are given a real function $K(\Phi^i, \overline{\Phi}^{\bar{\imath}})$. Then because the Berezinian line bundle of $\mathbf{\Sigma}$ is trivial, we can integrate $K(\Phi^i, \overline{\Phi}^{\bar{\imath}})$ over $\mathbf{\Sigma}$ to define the action of a field theory
\begin{equation}
I_{\text{B}} := \int_{\mathbf{\Sigma}} \mathcal{D}(z,\bar{z}|\theta^{\mp},\bar{\theta}^{\mp}) K(\Phi^i, \overline{\Phi}^{\bar{\imath}}),
\end{equation}
where $\mathcal{D}(z,\bar{z}|\theta^{\mp},\bar{\theta}^{\mp})$ is a section of the (trivial) Berezinian line bundle of $\mathbf{\Sigma}$. This is the action for the usual topological A model and B model, depending on whether we are using $\mathbf{\Sigma}_A$ or $\mathbf{\Sigma}_B$. The lowest components, namely $(\phi^i, \bar{\phi}^{\bar{\imath}})$, are interpreted as local coordinates on some target space $M$, which is usually a K\"{a}hler manifold (it has to be a Calabi-Yau manifold in the B model).

Aesthetically speaking, the above construction is somewhat unsatisfying as it is asymmetric: the worldsheet $\mathbf{\Sigma}$ is a supermanifold, but the target space $M$ is only an ordinary manifold. Our goal here is to generalize the above construction to supermanifolds. In other words, we would like to construct a topological sigma model such that its target space is a supermanifold $\mathbf{M}$. The above usual case can then be consider as a special case, in which the number of odd dimensions of $\mathbf{M}$ is 0.

To achieve this, we will proceed with a local approach. In particular, we will start off by considering odd complex functions on $\mathbf{\Sigma}$. Let $\Psi: \mathbf{\Sigma} \rightarrow \mathbb{C}^{0|1}$ be such an odd function. We call $\Psi$ an odd chiral superfield\footnote{This should not be confused with the usual (0,2) fermi superfields in the literature. See also in \cite{sethi} for some early discussions.} if 
\begin{equation}
D_- \Psi = \overline{D}_- \Psi = 0.
\end{equation}
Analogously, we call a function $\overline{\Psi}: \mathbf{\Sigma} \rightarrow \mathbb{C}^{0|1}$ an odd antichiral superfield if 
\begin{equation}
D_+ \overline{\Psi} = \overline{D}_+ \overline{\Psi} = 0.
\end{equation}

In terms of local component fields, we can make the following expansion:
\begin{equation}
\begin{split}
\Psi  =& \ \psi + \theta^+ \rho + \bar{\theta}^+ \xi  + \theta^+ \bar{\theta}^+ G - \theta^-\theta^+ \partial_z \psi - \bar{\theta}^- \bar{\theta}^+ \partial_{\bar{z}}\psi \\
& - \theta^+ \bar{\theta}^- \bar{\theta}^+ \partial_{\bar{z}} \rho - \theta^- \theta^+ \bar{\theta}^+ \partial_z \xi + \theta^- \theta^+ \bar{\theta}^- \bar{\theta}^+ \partial_z \partial_{\bar{z}} \psi, \\
\overline{\Psi} =& \ \bar{\psi} + \theta^- \bar{\rho} + \bar{\theta}^- \bar{\xi} +  \theta^- \bar{\theta}^- \bar{G}  + \theta^-\theta^+ \partial_{z} \bar{\psi} + \bar{\theta}^- \bar{\theta}^+ \partial_{\bar{z}}\bar{\psi} \\
&+ \theta^- \bar{\theta}^- \bar{\theta}^+ \partial_{\bar{z}} \bar{\rho} + \theta^- \theta^+ \bar{\theta}^- \partial_{z}\bar{\xi} + \theta^- \theta^+ \bar{\theta}^- \bar{\theta}^+ \partial_z \partial_{\bar{z}} \bar{\psi}.
\end{split}
\end{equation}
Once again, these component fields all have various geometric meanings depending whether we are dealing with $\mathbf{\Sigma}_A$ or $\mathbf{\Sigma}_B$. For now, let's focus on the lowest components $(\psi, \bar{\psi})$. Naturally, these are local coordinates parametrizing the targets space $\mathbb{C}^{0|1}$.

What's missing now is a way to patch together these odd local data together, in addition to the even local data coming from chiral superfields. A global justification can be found in appendix \ref{maps}. Briefly speaking, the definition of morphism between supermanifolds (as supercommutative locally ringed spaces) require the existence of both even and odd local functions, represented by superfields like $\Phi$ and $\Psi$. Let's say we have $n$ chiral superfields $\Phi^i$ and antichiral superfields $\overline{\Phi}^{\bar{\imath}}$, together with $m$ odd chiral superfields $\Psi^{\alpha}$ and odd antichiral superfields $\overline{\Psi}^{\bar{\alpha}}$. In addition, we need a real function $K(\Phi^i, \overline{\Phi}^{\bar{\imath}} | \Psi^{\alpha}, \overline{\Psi}^{\bar{\alpha}})$. Then we define the action of our topological sigma model as
\begin{equation} \label{action}
I := \int_{\mathbf{\Sigma}} \mathcal{D}(z,\bar{z}|\theta^{\mp},\bar{\theta}^{\mp}) K(\Phi^i, \overline{\Phi}^{\bar{\imath}} | \Psi^{\alpha}, \overline{\Psi}^{\bar{\alpha}}).
\end{equation}
We interpret the lowest components, namely $(\phi^i, \bar{\phi}^{\bar{\imath}} | \psi^{\alpha}, \bar{\psi}^{\bar{\alpha}})$, as local coordinates parametrizing a supermanifold $\mathbf{M}$ of complex dimension $n|m$. We can think of our topological sigma model defined by $I$ as a quantum field theory governing maps from $\mathbf{\Sigma}$ to $\mathbf{M}$, both of which are supermanifolds. In particular, our target supermanifold $\mathbf{M}$ is a super K\"{a}hler manifold with super K\"{a}hler potential $K(\Phi^i, \overline{\Phi}^{\bar{\imath}} | \Psi^{\alpha}, \overline{\Psi}^{\bar{\alpha}})$.

It is not enough to just have an action to define a topological sigma model. In addition, we need to define suitable physical observables so that we can compute their correlation functions. We will make this concrete separately for the A model and B model later in this paper. For now, let's say we have properly defined a set of physical observables $\mathcal{O}_a$. The goal of studying a topological field theory is to find the meaning of correlation functions defined by path integral
\begin{equation}
\braket{\prod_a \mathcal{O}_a} := \int \mathcal{D}[\Phi, \overline{\Phi}, \Psi, \overline{\Psi}] \ e^{-I} \prod_a \mathcal{O}_a.
\end{equation}
In general, this is notoriously difficult to compute (it is in fact not even mathematically well defined). However, one of the nicest properties of topological field theories is that their path integrals can be computed (at least formally) exactly.

\section{The A Model} \label{A}

When we take the worldsheet in the action (\ref{action}) to be $\mathbf{\Sigma}_A$, we call the resulting theory the A model. In this case, the component fields of chiral superfields have the following geometric meanings:
\begin{equation}
\begin{split}
\chi^i &\in \Gamma(\phi^*T^{1,0}M), \ \ \ \ \lambda^i \in \Gamma(\overline{K}_{\Sigma} \otimes (\phi^*T^{0,1}M)^{\vee}), \\
\bar{\chi}^{\bar{\imath}} &\in \Gamma(K_{\Sigma} \otimes (\phi^*T^{1,0}M)^{\vee}), \ \ \ \ \bar{\lambda}^{\bar{\imath}} \in \Gamma(\phi^*T^{0,1}M),
\end{split}
\end{equation}
where $\Sigma = {\mathbf{\Sigma}_A}_{\text{red}}$, $M = \mathbf{M}_{\text{red}}$, and $\phi: \Sigma \rightarrow M$ is the underlying map of the map between $\mathbf{\Sigma}$ and $\mathbf{M}$ that defines our sigma model. Note that $\chi^i$ and $\bar{\lambda}^{\bar{\imath}}$ are complex conjugate to each other, while $\bar{\chi}^{\bar{\imath}}$ and $\lambda^i$ are complex conjugate to each other.

Similarly, the component fields of odd chiral superfields have the following geometric meanings:
\begin{equation}
\begin{split}
\rho^{\alpha} &\in \Gamma(\phi^*T_-^{1,0}\mathbf{M}), \ \ \ \ \xi^{\alpha} \in \Gamma(\overline{K}_{\Sigma} \otimes (\phi^*T_-^{0,1}\mathbf{M})^{\vee}), \\
\bar{\rho}^{\bar{\alpha}} &\in \Gamma(K_{\Sigma} \otimes (\phi^*T_-^{1,0}\mathbf{M})^{\vee}), \ \ \ \ \bar{\xi}^{\bar{\alpha}} \in \Gamma(\phi^*T_-^{0,1}\mathbf{M}),
\end{split}
\end{equation}
where we have complexified the odd tangent bundle $T_-\mathbf{M}$ of $\mathbf{M}$. Again, note that $\rho^{\alpha}$ is the complex conjugate of $\bar{\xi}^{\bar{\alpha}}$, while $\bar{\rho}^{\bar{\alpha}}$ is the complex conjugate of $\xi^{\alpha}$.

\subsection{Localization}

In \cite{witten-tft}, Witten proposed a fixed point theorem for any quantum field theory with a fermionic symmetry. It is usually called supersymmetric localization in the literature. Here we will apply supersymmetric localization to our topological sigma models.

For the A model, let's define the generator of our fermionic symmetry to be
\begin{equation}
\mathcal{Q}_A := \overline{\mathcal{Q}}_- + \mathcal{Q}_+.
\end{equation}
This is the linear combination of the two scalar nilpotent supersymmetry charges, namely $\overline{\mathcal{Q}}_-$ and $\mathcal{Q}_+$, on $\mathbf{\Sigma}_A$. Then supersymmetric localization tells us that the path integral of our A model localizes to an integral over the fixed points of $\mathcal{Q}_A$.

So we are looking for the solutions to the following equations:
\begin{equation}
\mathcal{Q}_A \Phi^i = \mathcal{Q}_A \overline{\Phi}^{\bar{\imath}} = \mathcal{Q}_A \Psi^{\alpha} = \mathcal{Q}_A \overline{\Psi}^{\bar{\alpha}} = 0.
\end{equation}
From appendix \ref{susy-transf}, we know that, in terms of component fields, these become
\begin{equation} \label{a-locus}
\begin{split}
&\partial_z \bar{\phi}^{\bar{\imath}} = \partial_{\bar{z}} \phi^i = 0, \\
&\chi^i = F^i = \bar{\lambda}^{\bar{\imath}} = \bar{F}^{\bar{\imath}} = 0, \\
&\partial_z \bar{\psi}^{\bar{\alpha}} = \partial_{\bar{z}} \psi^{\alpha} = 0, \\
&\rho^{\alpha} = G^{\alpha} = \bar{\xi}^{\bar{\alpha}} = \bar{G}^{\bar{\alpha}} = 0.
\end{split} 
\end{equation}
This set of solutions is usually called localization locus, or BPS locus. The path integral of our theory localizes to this set of solutions.

Part of these solutions should be immediately recognizable, namely the first two lines of (\ref{a-locus}). This is the semiclassical locus for the ordinary A model with bosonic target spaces; usually it is called the moduli space of worldsheet instantons. Mathematically, it is the space of pseudo-holomorphic maps from $\Sigma$ to $M$.

The entire solution set of (\ref{a-locus}) is somewhat more difficult to grasp. The lowest component fields of odd chiral superfields, namely $\psi^{\alpha}$ and $\bar{\psi}^{\bar{\alpha}}$, are also ``pseudo-holomorphic''. Note that $\psi^{\alpha}$ and $\bar{\psi}^{\bar{\alpha}}$ are anti-commuting fields with spin 0. Therefore, we see that the path integral of our A model localizes to an integral over a supermanifold, which we denote as $\mathfrak{M}$.

It is a bit difficult to understand what $\mathfrak{M}$ is. By definition, it should be called the supermoduli space of ``superinstantons'', which are holomorphic maps from $\mathbf{\Sigma}$ to $\mathbf{M}$. The reduced space of $\mathfrak{M}$ is the usual worlsheet instanton moduli space. Note $\mathfrak{M}$ in general may have multiple disconnected components, so in general integrals over it should be expressed as a sum over integrals over each individual component. Here we will not go into those details to keep notations simple.

There is, however, one more layer of subtleties: we have to consider possible integration over the zero modes of various fields in our A model. More precisely, we have to consider the integral over the zero modes of $\xi^i$, $\bar{\lambda}^{\bar{\imath}}$, $\rho^{\alpha}$ and $\bar{\xi}^{\bar{\alpha}}$. We turn to this issue now.

\subsection{Ghost Number and Picture Number}

In the ordinary A model, there is a classical ghost number symmetry whose anomaly provides a selection rule for computing correlation functions. Similarly, there are two independent classical ghost number symmetries, $U(1)_{\Phi}$ and $U(1)_{\Psi}$, in our A model on supermanifold $\mathbf{M}$. In particular, for $U(1)_{\Phi}$ we assign 0 as the ghost number of $(\phi^i, \bar{\phi}^{\bar{\imath}})$, 1 as the ghost number of $(\chi^i, \bar{\lambda}^{\bar{\imath}})$, and -1 as the ghost number of $(\bar{\chi}^{\bar{\imath}}, \lambda^i)$. Analogously, for $U(1)_{\Psi}$ we assign ghost number 0 to $(\psi^{\alpha}, \bar{\psi}^{\bar{\alpha}})$, ghost number 1 to $(\rho^{\alpha}, \bar{\xi}^{\bar{\alpha}})$, and ghost number -1 to $(\bar{\psi}^{\bar{\alpha}}, \xi^{\alpha})$.

Again, similar to the ordinary A mode, our ghost number symmetries suffer from quantum anomalies in general. This is a result of integrating over the zero modes of various fields in our theory. Let's first look at the anomaly of $U(1)_{\Phi}$. Here the analysis is exactly the same as in the usual A model. Let $a$ be the number of $\chi$ zero modes, and $b$ be the number of $\bar{\chi}^{\bar{\imath}}$ zero modes\footnote{The number of zero modes of a section of a vector bundle $V$ is given by the dimension of the zero-th cohomology group $H^0(V)$.}. Then using the Hirzebruch-Riemann-Roch theorem, we can compute 
\begin{equation}
w := a - b = n(1 - g) + \int_{\Sigma} c_1(\phi^*TM),
\end{equation}
where $n$ is the complex dimension of $M$, $g$ is the genus of $\Sigma$, and $c_1(\phi^*TM)$ is the first Chern class of $\phi^*TM$ over $\Sigma$. Therefore, if we want to get a nonzero result from computing a correlation function $\braket{\prod_k \mathcal{O}_k}$ of some observables $\mathcal{O}_k$, the $U(1)_{\Phi}$ ghost number of $\prod_k \mathcal{O}_k$ must be the same as $w$, due to the natural of Grassmann integrals.

Things are quite different for $U(1)_{\Psi}$. It is important to keep in mind that $(\rho^{\alpha}, \bar{\xi}^{\bar{\alpha}})$ and $(\bar{\psi}^{\bar{\alpha}}, \xi^{\alpha})$ are commuting (bosonic) fields. As such, the integral over their zero modes is different from Grassmann integrals. The key observation is that their integral are very similar to the integral of the $\beta-\gamma$ system in superstring theory (see \cite{witten-spt} for an extensive discussion). The upshot is that we need delta functions (or more precisely distributions) of these fields. 

Let $t$ be the number of $\rho$ zero modes, and $s$ be the number of $\bar{\rho}$ zero modes. Then once again Hirzebruch-Riemann-Roch theorem tells us that the $U(1)_{\Psi}$ anomaly is associated with 
\begin{equation}
u := t - s = m (1 - g) + \int_{\Sigma} c_1(\phi^*T_-\mathbf{M}),
\end{equation}
where $m = \text{rank}(T_-\mathbf{M})$, with $T_-\mathbf{M}$ being the odd tangent bundle of $\mathbf{M}$. In contrast to the above case of $U(1)_{\Phi}$ anomaly, here we need the total $U(1)_{\Psi}$ ghost number of $\braket{\prod_k \mathcal{O}_k}$ to be $u$ so that the integral over these bosonic zero modes is well defined. If not, then we get a divergent, not vanishing, result.

This number $u$ is essential when we get to the next section. We will see that physical observables in our A model comes naturally with a label called picture number. In order for us to get a well defined integral over the localization locus $\mathfrak{M}$, we need, in generic cases when there are no $\bar{\rho}$ zero modes, the total picture number of observables in a correlation function to be $-u$. 

We call the pair $w|u$ the virtual dimension of our localization locus $\mathfrak{M}$. The reason for this is similar to the case of ordinary A model. Note the zero-mode equation of $\chi^i$ and $\rho^{\alpha}$ are linearized versions of $\partial_z \phi^i = 0$ and $\partial_{\bar{z}} \psi^{\alpha} = 0$, respectively. Hence the space of $\chi$ and $\rho$ zero modes are precisely $T_+\mathfrak{M}$ and $T_-\mathfrak{M}$, respectively, where $+$ and $-$ denote even and odd tangent bundles. Therefore, if we have a case in which $b = s = 0$, then the dimension of $\mathfrak{M}$ is precisely $w|u$.

\subsection{Observables and Correlation Functions}

The fundamental property of a cohomological topological field theory is to define its physical Hilbert space as the cohomology of some suitable nilpotent operator. Here we have our natural fermionic operator $\mathcal{Q}_A$, so we define the physical observables as elements of the cohomology of $\mathcal{Q}_A$ (or more precisely, the cohomology of the representation of $\mathcal{Q}_A$ on the space of fields). 

From the last section, a naive guess is that such local operators are of the form (in terms of component fields):
\begin{equation}
W_{i_1...i_p,\bar{\imath}_1...\bar{\imath}_q | \alpha_1...\alpha_k, \bar{\alpha}_1...\bar{\alpha}_{l}}(\phi^i, \bar{\phi}^{\bar{\imath}}, \psi, \bar{\psi}^{\bar{\alpha}}) \chi^{i_1} \cdots \chi^{i_p} \bar{\lambda}^{\bar{\imath}_1} \cdots \bar{\lambda}^{\bar{\imath}_q} \rho^{\alpha_1} \cdots \rho^{\alpha_k} \bar{\xi}^{\bar{\alpha}_1} \cdots \bar{\xi}^{\bar{\alpha}_l},
\end{equation}
where $W_{i_1...i_p,\bar{\imath}_1...\bar{\imath}_q | \alpha_1...\alpha_k, \bar{\alpha}_1...\bar{\alpha}_{l}}(\phi^i, \bar{\phi}^{\bar{\imath}}, \psi, \bar{\psi}^{\bar{\alpha}})$ is a function of $\phi^i, \bar{\phi}^{\bar{\imath}}, \psi, \bar{\psi}^{\bar{\alpha}}$ only.

As in the case of ordinary A model, such an operator corresponds to a differential form on $\mathbf{M}$. There is in fact an isomorphism between the de Rham cohomology of $\mathbf{M}$ and the cohomology of $\mathcal{Q}_A$, given by
\begin{equation} \label{to-de-rham}
\begin{split}
&\chi^i \rightarrow d\phi^i, \ \ \ \ \ \bar{\lambda}^{\bar{\imath}} \rightarrow d\bar{\phi}^{\bar{\imath}} \\
&\rho^{\alpha} \rightarrow d\psi^{\alpha}, \ \ \ \ \bar{\xi}^{\bar{\alpha}} \rightarrow d\bar{\psi}^{\bar{\alpha}}
\end{split}
\end{equation}

However, unlike the ordinary A model, our localization locus is a supermanifold $\mathfrak{M}$. Differential forms are not suitable to be integrated over on supermanifolds; instead we need integral forms \cite{witten-sm}. Therefore, we would like to define our physical observables as\footnote{See \cite{sethi} from some early discussions.}:
\begin{equation}
\mathcal{O}_W := W_{i_1...i_p,\bar{\imath}_1...\bar{\imath}_q | \alpha_1...\alpha_k, \bar{\alpha}_1...\bar{\alpha}_{l}}(\phi^i, \bar{\phi}^{\bar{\imath}}, \psi, \bar{\psi}^{\bar{\alpha}}) \chi^{i_1} \cdots \chi^{i_p} \bar{\lambda}^{\bar{\imath}_1} \cdots \bar{\lambda}^{\bar{\imath}_q} \delta(\rho^{\alpha_1}) \cdots \delta(\rho^{\alpha_k}) \delta(\bar{\xi}^{\bar{\alpha}_1}) \cdots \delta(\bar{\xi}^{\bar{\alpha}_l}),
\end{equation}
which is manifestly invariant under $\mathcal{Q}_A$. $\mathcal{O}_W$ is said to have superdegree $p,q|k,l$, with picture number $-k-l$.

Under the above isomorphism (\ref{to-de-rham}), our physical observables like $\mathcal{O}_W$ are in one-to-one correspondence with pseudoforms\footnote{We use the term pseudoform as defined by Witten in \cite{witten-sm}. Pseudoforms sometimes have a larger meaning in the literature; the forms we are considering can always be obtained from general pseudoforms through the Baranov-Schwartz transformation \cite{baranov-schwartz}.} on $\mathfrak{M}$. For example, $\mathcal{O}_W$ corresponds to a pseudoform of superdegree $p,q|k,l$. Therefore, the BRST cohomology of our A model is isomorphic to the cohomology of pseudoforms on $\mathbf{M}$.

In order to meaningfully integrate over $\mathfrak{M}$, we need the total superdgree to be of the dimension of $\mathfrak{M}$ in generic situations (i.e. when there are no $\bar{\chi}$ and $\bar{\rho}$ zero modes). More generally, we need the following selection rules:
\begin{equation} \label{A-select}
\begin{split}
\sum_a p_a &= \sum_a q_a = n(1 - g) + \int_{\Sigma} c_1(\phi^*TM), \\
\sum_a k_a &= \sum_a l_a = m (1 - g) + \int_{\Sigma} c_1(\phi^*T_-\mathbf{M}),
\end{split}
\end{equation}
for a collection of physical operators $\mathcal{O}_{W_a}$, where the notations on the right hand side are from the last section. In other words, we need to obtain an integral form on $\mathfrak{M}$ with top superdegree from suitable operator insertions.

Exactly analogous to the ordinary A model, the one-loop determinant our our general A model is simply 1. This is because, analogous to the case of ordinary A model, fields in our A model pair up as complex conjugates, and the existence of BRST symmetry ensures cancellation between bosonic and fermionic determinants.

Therefore, combining every ingredients, we see that our correlation function becomes (in generic situations)
\begin{equation}
\braket{\prod_a \mathcal{O}_{W_a}} = \int_{\mathfrak{M}} \ \prod_a V_a,
\end{equation}
where
\begin{equation}
V_a := W_a d\phi^{i_1} \cdots d\phi^{i_{p_a}} d\bar{\phi}^{\bar{\imath}_1} \cdots d\bar{\phi}^{\bar{\imath}_{i_{q_a}}} \delta(d\psi^{\alpha_1}) \cdots \delta(d\psi^{\alpha_{k_a}}) \delta(d\bar{\psi}^{\bar{\alpha}_1}) \cdots \delta(d\bar{\psi}^{\bar{\alpha}_{l_{a}}}),
\end{equation}
is the corresponding pseudoform on $\mathfrak{M}$ (pulled back from $\mathbf{M}$) to $\mathcal{O}_{W_a}$.

In non-generic cases, things are somewhat more complicated. The dimension of $\mathfrak{M}$ in these cases are not the same as its virtual dimension. Therefore, aside from utilizing all operator insertions, we also need to consider the contribution from integrating over $\bar{\chi}$ and $\bar{\rho}$ zero modes. In the case of the ordinary A model, this amounts to inserting a factor of the Euler class of the obstruction bundle (the bundle of $\bar{\chi}$ zero modes).

In our case, we have both $\bar{\chi}$ and $\bar{\rho}$ zero modes. Therefore, we obtain an super vector bundle over $\mathfrak{M}$. Let's call this super vector bundle $\mathbf{V}$, which has rank $b|s$ (notation from the last section). Applying the standard argument from ordinary A model using four-fermi interaction terms, we see that the integral over $\bar{\chi}$ zero modes provides a differential form which can be thought of as the Euler class of $\mathbf{V}_+$ (the bosonic subbundle of $\mathbf{V}$). Completely analogous to this, we can integrate over $\bar{\rho}$ zero modes and obtain a delta function (using the similarity to the $\beta-\gamma$ system studied in \cite{witten-spt, mine-ts}) of $\mathbf{V}_-$. Combining these two pieces together, we obtain a pseudoform, which we call $e(\mathbf{V})$, of superdegree $2b|2s$. Then general correlation functions can be expressed as
\begin{equation}
\braket{\prod_a \mathcal{O}_{W_a}} = \int_{\mathfrak{M}} \ e(\mathbf{V})\prod_a V_a. 
\end{equation}
Note that the appearance of $e(\mathbf{V})$, together with our selection rules (\ref{A-select}), makes this formula a well defined expression.

\subsection{If $\mathfrak{M}$ is Split}

One of the major issues of performing integrations over general supermanifolds is how to handle integrals over odd coordinates. In general, if a supermanifold is not projected (hence not split), we cannot consistently integrate over odd coordinates on it and reduce to an integral over its reduced space. This is a major driving force for recent works on superstring perturbation theory \cite{witten-spt} and topological string theory \cite{mine-ts}.

When a supermanifold is indeed split, which by definition means that globally it is a vector bundle over its reduced space, then we are allowed to integrate over the odd directions first without any issues. The correct procedure to do this, as first developed in superstring perturbation theory, is to use the so called picture-changing operator \cite{witten-spt}. 

Picture-changing operator has a purely geometric interpretation, as discovered in \cite{belopolsky-picture}. It is an operation that consistently map pseudoforms of superdegree $n|m$ to pseudoforms of superdegree $n|m-1$, or in other words increase picture number by 1 (hence its name). Abstractly, it is defined to be
\begin{equation}
\Gamma_{\nu} := \frac{1}{2} \left( \delta(\mathbf{i}_{\nu}) \mathfrak{L}_{\nu} + \mathfrak{L}_{\nu} \delta(\mathbf{i}_{\nu}) \right),
\end{equation}
where $\nu$ is an odd vector field, $\delta(\mathbf{i}_{\nu})$ is defined in appendix \ref{forms}, and $\mathfrak{L}_{\nu}$ is the Lie derivative
\begin{equation}
\mathfrak{L}_{\nu} := \text{d} \mathbf{i}_{\nu} + \mathbf{i}_{\nu} \text{d}.
\end{equation}

The procedure of using picture-changing operators to consistently integrate out odd coordinates is the following: when we integrate a particular odd direction specified by $\nu$, we replace the integrand, say $\omega$, by $\Gamma_{\nu}(\omega)$. By doing so repeatedly, which is only strictly allowed if our supermanifold is split, we can reduce an integral over the entire supermanifold to an integral over its reduced space.

Let's try to apply this to our current situation. If the supermoduli space $\mathfrak{M}$ of superinstantons is split, then we can rewrite correlation functions using picture-changing operators. More explicitly, let $N|M$ be the complex dimension of $\mathfrak{M}$, then
\begin{equation}
\int_{\mathfrak{M}} \prod_a V_a = \int_{\mathcal{M}} \Gamma_{\alpha}^{2M}\left(e(\mathbf{V})\prod_a V_a\right)
\end{equation}
where $\mathcal{M}$ is the usual instanton moduli space, and $\Gamma_{\alpha}^{2M}$ denotes applying picture-changing operator enough times to remove integrals over all odd coordinates.

To sum up, when our supermoduli space $\mathfrak{M}$ is split, we can use picture-changing operator to remove integrals over odd coordinates. In the end, we end up with an integral over the ordinary instanton moduli spaces. This might provide a relationship between correlation functions in our general A model and correlation functions in ordinary A model.\footnote{See \cite{schwarz} for some related discussion.}

\section{The B Model} \label{B}

Now we turn to the B model, which is obtained by taking the worldsheet to be $\mathbf{\Sigma}_B$. In this case, the geometric meanings of component fields in chiral superfields become
\begin{equation}
\begin{split}
\chi^i &\in \Gamma(K_{\Sigma} \otimes \phi^*T^{1,0}M), \ \ \lambda^i \in \Gamma(\overline{K}_{\Sigma} \otimes (\phi^*T^{0,1}M)^{\vee}), \\
\bar{\chi}^{\bar{\imath}} &\in \Gamma((\phi^*T^{1,0}M)^{\vee}), \ \ \ \ \ \ \ \bar{\lambda}^{\bar{\imath}} \in \Gamma(\phi^*T^{0,1}M),
\end{split}
\end{equation}

On the other hand, component fields in odd chiral superfields become
\begin{equation}
\begin{split}
\rho^{\alpha} &\in \Gamma(K_{\Sigma} \otimes \phi^*T_-^{1,0}\mathbf{M}), \ \ \xi^{\alpha} \in \Gamma(\overline{K}_{\Sigma} \otimes (\phi^*T_-^{0,1}\mathbf{M})^{\vee}), \\
\bar{\rho}^{\bar{\alpha}} &\in \Gamma((\phi^*T_-^{1,0}\mathbf{M})^{\vee}), \ \ \ \ \ \ \ \bar{\xi}^{\bar{\alpha}} \in \Gamma(\phi^*T_-^{0,1}\mathbf{M}),
\end{split}
\end{equation}

In \cite{witten-twistor}, the special case in which the target supermanifold is a complex superprojective space $\mathbb{CP}^{n|m}$ was studied to some extent. Here in this paper, we will establish general properties of the B model on general supermanifolds.

\subsection{Localization and Anomalies}

As in the case of the A model, we would like to apply supersymmetric localization to our B model. In this case, we define our nilpotent BRST operator to be
\begin{equation}
\mathcal{Q}_B := \overline{\mathcal{Q}}_- + \mathcal{Q}_-
\end{equation}
based on the natural structure of $\mathbf{\Sigma}_B$.

With this definition of BRST operator, the fixed points we are looking for are solutions of 
\begin{equation}
\mathcal{Q}_B \Phi^i = \mathcal{Q}_B \overline{\Phi}^{\bar{\imath}} = \mathcal{Q}_B \Psi^{\alpha} = \mathcal{Q}_B \overline{\Psi}^{\bar{\alpha}} = 0.
\end{equation}
Based on the results of appendix \ref{susy-transf}, we obtain the equivalent equations in terms of component fields:
\begin{equation} \label{b-locus}
\begin{split}
&\partial_z \phi^i = \partial_{\bar{z}} \phi^i = 0, \\
&\partial_{\bar{z}} \chi^i = \partial_z \lambda^i = \bar{\chi}^{\bar{\imath}} = \bar{\lambda}^{\bar{\imath}} = \bar{F}^{\bar{\imath}} = 0, \\
&\partial_z \psi^{\alpha} = \partial_{\bar{z}} \psi^{\alpha} = 0, \\
&\partial_{\bar{z}} \rho^{\alpha} = \partial_z \xi^{\alpha} = \bar{\rho}^{\bar{\alpha}} = \bar{\xi}^{\bar{\alpha}} = \bar{G}^{\bar{\alpha}} = 0.
\end{split}
\end{equation}\
The solutions to these equations is the localization locus of our B model. The path integral of our B model localizes to this set.

Once again, the first two lines of (\ref{b-locus}) should be familiar, as they are characterizing the BPS locus of the ordinary B model on bosonic target spaces. More precisely, the ordinary B model localizes to the space of constant maps from worldsheet to target space.

The entire set of solutions to (\ref{b-locus}) is not that difficult to understand; it is very much simpler than the A model case. Here the full localization locus is once again a supermanifold, which is the space of constant maps from worldsheet $\mathbf{\Sigma}_B$ to target space $\mathbf{M}$. Therefore it is simply a copy of $\mathbf{M}$. This is in accordance with the ordinary B model on bosonic target spaces.

Analogous to the A model case, we need to worry about possible zero modes from various fields which contribute to anomalies. The situation is, however, very different to the A model case. Here, the fields that contribute to anomalies don't pair up as complex conjugate to each other as in the A model. As such, we obtain some complex fermionic determinants which must be handled by suitable anomaly cancellation conditions without involving any operator insertions. In the ordinary B model with even target spaces, this is achieved by demanding the first Chern class of the target space to vanish. Here we have a similar situation, and anomalies are cancelled by demanding that $\mathbf{M}$ must be a Calabi-Yau supermanifold\footnote{See \cite{witten-twistor} for some relevant discussion.}. 

On top of this Calabi-Yau condition, we need some more selection rules so that we can obtain a non-vanishing and well-defined result when computing correlation functions $\braket{\prod_a \mathcal{O}_a}$. The details will be given in the next section after we have identified suitable physical observables in our B model.

\subsection{Observables and Correlation Functions}

We are now naturally looking for the cohomology of our operator $\mathcal{Q}_B$. It is traditional to define some new fields:
\begin{equation}
\begin{split}
\eta^{\bar{\imath}} &:= \bar{\chi}^{\bar{\imath}} + \bar{\lambda}^{\bar{\imath}}, \\
\theta_i &:= g_{i\bar{\imath}} (\bar{\chi}^{\bar{\imath}} - \bar{\lambda}^{\bar{\imath}}), \\
\kappa^{\bar{\alpha}} &:= \bar{\rho}^{\bar{\alpha}} + \bar{\xi}^{\bar{\alpha}}, \\
\zeta_{\alpha} &:= g_{\alpha\bar{\alpha}} (\bar{\rho}^{\bar{\alpha}} - \bar{\xi}^{\bar{\alpha}}),
\end{split}
\end{equation}
where $g_{i\bar{\imath}}$ and $g_{\alpha\bar{\alpha}}$ are parts of the supermetric tensor on the target supermanifold $\mathbf{M}$ defined by $K$. Note these fields are all scalars on the worldsheet $\Sigma$.

In the ordinary B model with even target space $M$, the physical observables are of the form
\begin{equation}
W(\phi,\bar{\phi})_{\bar{\imath}_1...\bar{\imath}_{p}}^{i_1...i_q} \eta^{\bar{\imath}_1} \cdots \eta^{\bar{\imath}_p} \theta_{i_1} \cdots \theta_{i_q}
\end{equation}
Under the maps $\eta^{\bar{\imath}} \rightarrow d\bar{\phi}^{\bar{\imath}}$, $\theta_{i} \rightarrow \partial_{\phi^i}$, these can be mapped to elements of the sheaf cohomology group $H^p(M, \wedge^q T^{1,0}M)$. There, the BRST operator is identified as the $\bar{\partial}$ operator on $M$.

Here, we have a target supermanifold $\mathbf{M}$. Based on our previous experience with the A model, we postulate that physical observables of our B model are of the form:
\begin{equation}
\mathcal{O}_W := W(\phi,\bar{\phi}|\psi,\bar{\psi})_{\bar{\imath}_1...\bar{\imath}_{p}|\bar{\alpha}_1...\bar{\alpha}_k}^{i_1...i_q|\alpha_1...\alpha_l} \eta^{\bar{\imath}_1} \cdots \eta^{\bar{\imath}_p} \theta_{i_1} \cdots \theta_{i_q} \delta(\kappa^{\bar{\alpha}_1}) \cdots \delta(\kappa^{\bar{\alpha}_k}) \delta(\zeta_{{\alpha_1}}) \cdots \delta(\zeta_{{\alpha}_l}),
\end{equation}
which are obviously closed under the action of our BRST operator $\mathcal{Q}_B$. 

Let's consider, in addition to the identifications in the ordinary B model listed above, maps $\kappa^{\bar{\alpha}} \rightarrow d\bar{\psi}^{\bar{\alpha}}$, $\zeta_{\alpha} \rightarrow \partial_{\psi^{\alpha}}$. Let $\mathcal{F}$ be the set of functions on $T^*\mathbf{M}$. Then our physical operators like $\mathcal{O}_W$ are in one-to-one correspondence with pseudoforms on $\mathbf{M}$ of purely anti-holomorphic superdegrees with values in $\mathcal{F}$, . 

From ghost numbers anomaly cancellations, we need to impose the following selection rules:
\begin{equation}
\begin{split}
\sum_a (p_a + q_a) &= 2n(1 - g), \\
\sum_a (k_a + l_a) &= 2m(1 - g),
\end{split}
\end{equation}
where $n|m$ is the complex dimension of $\mathbf{M}$, so that we can obtain a nonzero and well-defined correlation function $\braket{\prod_a \mathcal{O}_{W_a}}$. 

Let's now focus on the case where $g = 0$. It is important to keep in mind that $\mathbf{M}$ here is in fact a Calabi-Yau supermanifold. In particular, we are naturally provided with a supermetric tensor on $\mathbf{M}$. This allows ups, among other things, to map sections of $T^*\mathbf{M}$ to sections of $T\mathbf{M}$. So correlation functions in our B model are given by
\begin{equation}
\braket{\prod_a \mathcal{O}_a} = \int_{\mathbf{M}} f(\prod_a V_a),
\end{equation}
where
\begin{equation}
V_a := W_a d\bar{\phi}^{\bar{\imath}_1} \cdots d\bar{\phi}^{\bar{\imath}_{p_a}} \partial_{\phi_{i_1}} \cdots \partial_{\phi_{i_{q_a}}} \delta(d\bar{\psi}^{\bar{\alpha}_1}) \cdots \delta(d\bar{\psi}^{\bar{\alpha}_{k_a}}) \delta(\partial_{\psi_{\alpha_1}}) \cdots \delta(\partial_{\psi_{{\alpha}_{l_a}}}),
\end{equation}
and $f$ is the map from $\prod_a V_a$ to integral forms on $\mathbf{M}$, provided by the Calabi-Yau metric on $\mathbf{M}$.

Note that, just as in the ordinary B model, the correlation functions in our B model are purely classical, in the sense that there is no quantum corrections at all.

\section{Discussion} \label{discussion}

This paper primarily concerns topological sigma models governing maps from semirigid super Riemann surfaces and target supermanifolds, using techniques from supergeometry extensively. We have constructed suitable worldsheets and Lagrangians, defined appropriate physical observables and wrote down correlation functions using supersymmetric localization. Some general properties of these topological sigma models are analyzed. However, this paper has generated a lot more questions than answers.

One immediate question we should ask is: what are those correlation functions computing? Are they computing some topological invariants on supermanifolds, like in the case of ordinary topological A model? If so, can we find a mathematically rigorous treatment of this subject?

One possible way to achieve a better understanding of these correlation functions is to apply supersymmetric localization in its modern form. We might be able to explicitly compute these correlation functions in some cases and analyze their meanings. It remains to be seen whether this is possible.

Related to this, it is very natural for us to wonder if there is a generalization of mirror symmetry to general Calabi-Yau supermanifolds. There have been some related discussions to some level, see for example in \cite{sethi}. However, when lifted to the general playground we have been discussing, it is unclear how far we can push mirror symmetry. Hopefully, there is some sort of mirror symmetry relating the A model on a Calabi-Yau supermanifold and the B model on its ``mirror''.

One of the aims of this paper is to utilize the language of supergeometry throughout the process. However, there are still many places that are left behind. For example, physical operators are not expressed as anything in terms of supergeometry. As such, the current treatment in this paper is not completely satisfactory and it would be better if we could extend the usage of supergeometry further.

Finally, we could try to generalize even further. In the case of ordinary A model and B model, one can generalize to the so called ``half-twisted'' A/2 model and B/2 model. These theories concern more data: target spaces and suitable vector bundles over them. We could define analogous ``half-twisted'' sigma models in our general setup with supermanifolds, using ideas from heterotic superstring perturbation theory from \cite{witten-spt}.

Therefore, there are a lot more that are left to be done. We hope to solve some of these puzzles in future works.

\section*{Acknowledgement}
We would like to thank J. Distler and E. Sharpe for useful discussions. This work is supported by the National Science Foundation grant PHY-1316033.

\appendix

\section{Supermanifolds and Morphisms} \label{maps}

A supermanifold is generalization of the usual notion of manifold. Recall that an ordinary manifold can be though of as a pair $(M, \mathcal{O}_M)$, where $M$ is the underlying manifold, and $\mathcal{O}_M$ is its structure sheaf. 

Let $V \rightarrow M$ be a vector bundle over $M$. Let $\wedge^{\bullet} V^{\vee}$ be the wedge power of the dual bundle $V^{\vee}$. A split supermanifold $\mathbf{M}(M, V)$ is defined to be the pair $(M, \mathcal{O}_{\mathbf{M}})$, where $\mathcal{O}_{\mathbf{M}}$ is the sheaf of $\mathcal{O}_{M}$ valued sections of $\wedge^{\bullet} V^{\vee}$. Then we define a supermanifold $\mathbf{M}$ to be a supercommutative locally ringed space that is locally isomorphic to a split supermanifold $\mathbf{M}(M, V)$. 

Given a generic supermanifold $\mathbf{M}$, we can always recover a split supermanifold by modding out the nilpotents. More precisely, let $J$ be the ideal of $\mathcal{O}_{\mathbf{M}}$ consisting of all nilpotents. Then we can recover a pair $(M,V)$, where $\mathcal{O}_M = \mathcal{O}_{\mathbf{M}} / J$, and $V = J / J^2$. $(M, \mathcal{O}_M)$ is called the reduced space (or body) of $\mathbf{M}$, denoted as $\mathbf{M}_{\text{red}}$. 

Locally, the local coordinates on $\mathbf{M}_{\text{red}}$ provide local even coordinates on $\mathbf{M}$, while the local sections of $V^{\vee}$, with spin statistics reversed, provide local odd coordinates on $\mathbf{M}$. In other words, locally a supermanifold is simply isomorphic to $\mathbb{A}^{n|m}$, the usual affine superspace of dimension $n|m$. A generic supermanifold can also be defined by patching together these local data in a consistent way.

The tangent bundle $T\mathbf{M}$ of a supermanifold $\mathbf{M}$ is a sheaf of $\mathcal{O}_{\mathbf{M}}$ modules. When restricted to $\mathbf{M}_{\text{red}}$, it splits into two parts:
\begin{equation}
T_+\mathbf{M} : = T\mathbf{M}_{\text{red}}, \ \ \ T_-\mathbf{M} := V.
\end{equation}
$T_+\mathbf{M}$ and $T_-\mathbf{M}$ are called the even and odd tangent bundles of $\mathbf{M}$, respectively.

Morphisms between supermanifolds are defined similarly to those of ordinary locally ringed spaces. Let $\mathbf{X}$ and $\mathbf{Y}$ be two supermanifolds. A morphism between $\mathbf{X}$ and $\mathbf{Y}$ is a pair of maps $(f, \varphi)$, where $f: \mathbf{X}_{\text{red}} \rightarrow \mathbf{Y}_{\text{red}}$ is a (continuous, smooth, or analytic, as needed) map, while $\varphi: \mathcal{O}_{\mathbf{Y}} \rightarrow f_*\mathcal{O}_{\mathbf{X}}$ is a morphism between sheaves. Here $f_*\mathcal{O}_{\mathbf{X}}$ denote the direct image sheaf of $\mathcal{O}_{\mathbf{X}}$ under $f$. 

Locally speaking, given an open subset $U$ of $\mathbf{Y}$, the sheaf morphism $\varphi$ produce a homomorphism $\varphi_U: \mathcal{O}_{\mathbf{Y}}(U) \rightarrow \mathcal{O}_{\mathbf{X}}(f^{-1}U)$ of supercommutative rings, with some further consistent conditions that we ommit. Let $(y^i | \theta^{\alpha})$ be a set of local coordinates on $\mathbf{Y}$. Then it can be shown \cite{atenacci-debernardi-grassi-matessi} that $\varphi_U$ can be given by even functions $\Phi^i$ and odd functions $\Psi^{\alpha}$ on $f^{-1}U$, such that $\Phi^i = \varphi_U(y^i)$ and $\Psi^{\alpha} = \varphi_U(\theta^{\alpha})$. As their names suggest, $\Phi^i$ and $\Psi^{\alpha}$ are superfields used in the physics literature. This justifies the physics approach we have used to define sigma models governing maps between supermanifolds.

\section{Pseudoforms on Supermanifolds} \label{forms}

Let $\mathbf{M}$ be a supermanifold. Let $\mathit{\Pi} TM$ denotes the tangent bundle of $\mathbf{M}$ with statics reversed on each fiber. Consider the set of all generalized functions on $\mathit{\Pi} TM$. One can naturally define a Clifford multiplication and a Weyl multiplication on this set \cite{witten-sm}. A subset of these functions that are closed under these multiplications and form an algebra is called the space of pseudoforms. In fact, as discussed in \cite{witten-sm}, one would also like to impose a scale invariant property on these fors.

A special subclass of pseudoforms are simply differential forms. Locally, differential forms can be expressed as
\begin{equation}
f(t^1...t^n|\theta^1...\theta^m) dt^1 \cdots dt^p d\theta^1 \cdots d\theta^q,
\end{equation}
where $(t^1,...,t^n|\theta^1,...,\theta^m)$ are local coordinates on $\mathbf{M}$. Note that differential forms have polynomial dependence on one-forms like $dt^i$ and $d\theta^{\alpha}$. As such, there isn't any differential forms of top degree, which means that differential forms are not suitable to define integrals over supermanifolds.

Another special subclass of pseudoforms are called integral forms. Locally, these forms can be expressed, for example, as
\begin{equation}
f(t^1...t^n|\theta^1...\theta^m) dt^1 \cdots dt^p \delta(d\theta^1) \cdots \delta(d\theta^q).
\end{equation}
One defining property of integral forms is that they depend polynomially on odd one-forms like $dt^i$, but have distributional dependence on even one-forms like $d\theta^{\alpha}$. This fact indicates, among other things, that integral forms have a upper limit in terms of degree; integral forms of top degree can be integrated over $\mathbf{M}$.

In general, a pseudoform can have polynomial dependence on some one-forms but distributional dependence on others. One important property of pseudoforms is that they come with a grading, which is called superdegree. A pseudoform with superdegree $p|q$ scales as what one would expect under scaling of even and odd coordinates. We say that a pseudoform with superdegree $p|q$ has picture number $-q$.

Let's take a look at some natural operations on pseudoforms \cite{witten-sm, belopolsky-picture}. There is a natural exterior derivative
\begin{equation}
\text{d} := \sum_{I=1...n|1...m} \text{d}x^I \frac{\partial}{\partial x^I},
\end{equation}
where $x$ schematically denotes all the coordinates on $\mathbf{M}$, and $I$ is the index denoting all even and odd directions. This operator acts on the space of pseudoforms, mapping pseudoforms of superdegree $p|q$ to pseudoforms of superdegree $p+1|q$.

Let $V$ be a vector field on $\mathbf{M}$. Locally, we can express $V$ as
\begin{equation}
V = \sum_{I=1...n|1...m} V^I \frac{\partial}{\partial x^I}.
\end{equation}
Similar to differential geometry on ordinary manifolds, here we can define a contraction operator
\begin{equation}
\mathbf{i}_V := \sum_{I=1...n|1...m} V^I \frac{\partial}{\partial \text{d}x^I},
\end{equation}
which acts on pseudoforms, mapping forms of superdegree $p|q$ to forms of superdegree $p-1|q$. Based on this, one can define an operator $\delta(\mathbf{i}_V)$ that changes picture number. It is defined as
\begin{equation}
\begin{split}
&\delta(\mathbf{i}_V) := \mathbf{i}_V, \ \ \ \text{if} \ V \ \text{is even}, \\
& [\delta(\mathbf{i}_V) \omega] (x, \text{d}x) = \int [du] \omega(x, \text{d}x + uV), \ \text{if} \ V \ \text {is odd}.
\end{split}
\end{equation}
where $u$ is an even scaler and $\omega$ is a suitable pseudoform. Naturally, $\delta(\mathbf{i}_V)$ maps pseudoforms of superdegree $p|1$ to pseudoforms of superdegree $p|q-1$. 

Please refer to \cite{witten-sm, belopolsky-picture} for more details of properties of pseudoforms and operations on them.

\section{Supersymmetry Transformations} \label{susy-transf}

In physics, it is more common to write everything in terms of component fields rather than superfields. Traditionally, a quantum field theory is called supersymmetric if there is some fermionic transformations between these component fields such that the entire action is invariant under these transformations.

It is easy to derive those relevant supersymmetry transformations in topological sigma models we have studied in the main text. In this section we write them down to keep a record. We start by acting those supersymmetry charges on chiral superfields and odd chiral super fields:
\begin{equation}
\begin{split}
\mathcal{Q}_+ \Phi &= \chi + \bar{\theta}^+ F - \theta^- \theta^+ \partial_z \chi - \bar{\theta}^- \bar{\theta}^+ \partial_{\bar{z}} \chi - \theta^- \theta^+\bar{\theta}^+ \partial_z F + \theta^- \theta^+ \bar{\theta}^- \bar{\theta}^+ \partial_z \partial_{\bar{z}} \chi, \\
\mathcal{Q}_- \Phi &= -2 \theta^+ \partial_z \phi - 2 \theta^+ \bar{\theta}^+ \partial_z \lambda + 2 \theta^+ \bar{\theta}^- \bar{\theta}^+ \partial_z \partial_{\bar{z}} \phi, \\
\overline{\mathcal{Q}}_+ \Phi &= \lambda - \theta^+ F - \theta^- \theta^+ \partial_z \lambda - \bar{\theta}^- \bar{\theta}^+ \partial_{\bar{z}} \lambda - \theta^+ \bar{\theta}^- \bar{\theta}^+ \partial_{\bar{z}} F + \theta^- \theta^+ \bar{\theta}^- \bar{\theta}^+ \partial_z \partial_{\bar{z}} \lambda, \\
\overline{\mathcal{Q}}_- \Phi &= -2 \bar{\theta}^+ \partial_{\bar{z}} \phi + 2 \theta^+ \bar{\theta}^+ \partial_{\bar{z}} \chi + 2 \theta^- \theta^+ \bar{\theta}^+ \partial_z \partial_{\bar{z}} \phi.
\end{split}
\end{equation}

\begin{equation}
\begin{split}
\mathcal{Q}_+ \overline{\Phi} &= -2 \theta^- \partial_z \bar{\phi} - 2 \theta^- \bar{\theta}^- \partial_z \bar{\lambda} - 2 \theta^- \bar{\theta}^- \bar{\theta}^+ \partial_z \partial_{\bar{z}} \bar{\phi}, \\
\mathcal{Q}_- \overline{\Phi} &= \bar{\chi} + \bar{\theta}^- \bar{F} + \theta^- \theta^+ \partial_z \bar{\chi} + \bar{\theta}^- \bar{\theta}^+ \partial_{\bar{z}} \bar{\chi} - \theta^+ \theta^- \bar{\theta}^- \partial_z \bar{F} + \theta^- \theta^+ \bar{\theta}^- \bar{\theta}^+ \partial_z \partial_{\bar{z}} \bar{\chi}, \\
\overline{\mathcal{Q}}_+ \overline{\Phi} &= -2 \bar{\theta}^- \partial_{\bar{z}} \bar{\phi} + 2 \theta^- \bar{\theta}^- \partial_{\bar{z}} \bar{\chi} - 2 \theta^- \theta^+ \bar{\theta}^- \partial_z \partial_{\bar{z}} \bar{\phi}, \\ 
\overline{\mathcal{Q}}_- \overline{\Phi} &= \bar{\lambda} + \theta^- \bar{F} + \theta^- \theta^+ \partial_z \bar{\lambda} + \bar{\theta}^- \bar{\theta}^+ \partial_{\bar{z}} \bar{\lambda} - \theta^- \bar{\theta}^- \bar{\theta}^+ \partial_{\bar{z}} \bar{F} + \theta^- \theta^+ \bar{\theta}^- \bar{\theta}^+ \partial_z \partial_{\bar{z}} \bar{\lambda}.
\end{split}
\end{equation}

\begin{equation}
\begin{split}
\mathcal{Q}_+ \Psi &= \rho + \bar{\theta}^+ G - \theta^- \theta^+ \partial_z \rho - \bar{\theta}^- \bar{\theta}^+ \partial_{\bar{z}} \rho - \theta^- \theta^+\bar{\theta}^+ \partial_z G + \theta^- \theta^+ \bar{\theta}^- \bar{\theta}^+ \partial_z \partial_{\bar{z}} \rho, \\
\mathcal{Q}_- \Psi &= -2 \theta^+ \partial_z \psi - 2 \theta^+ \bar{\theta}^+ \partial_z \xi + 2 \theta^+ \bar{\theta}^- \bar{\theta}^+ \partial_z \partial_{\bar{z}} \psi, \\
\overline{\mathcal{Q}}_+ \Psi &= \xi - \theta^+ G - \theta^- \theta^+ \partial_z \xi - \bar{\theta}^- \bar{\theta}^+ \partial_{\bar{z}} \xi - \theta^+ \bar{\theta}^- \bar{\theta}^+ \partial_{\bar{z}} G + \theta^- \theta^+ \bar{\theta}^- \bar{\theta}^+ \partial_z \partial_{\bar{z}} \xi, \\
\overline{\mathcal{Q}}_- \Psi &= -2 \bar{\theta}^+ \partial_{\bar{z}} \psi + 2 \theta^+ \bar{\theta}^+ \partial_{\bar{z}} \rho + 2 \theta^- \theta^+ \bar{\theta}^+ \partial_z \partial_{\bar{z}} \psi. 
\end{split}
\end{equation}

\begin{equation}
\begin{split}
\mathcal{Q}_+ \overline{\Psi} &= -2 \theta^- \partial_z \bar{\psi} - 2 \theta^- \bar{\theta}^- \partial_z \bar{\xi} - 2 \theta^- \bar{\theta}^- \bar{\theta}^+ \partial_z \partial_{\bar{z}} \bar{\psi}, \\
\mathcal{Q}_- \overline{\Psi} &= \bar{\rho} + \bar{\theta}^- \bar{G} + \theta^- \theta^+ \partial_z \bar{\rho} + \bar{\theta}^- \bar{\theta}^+ \partial_{\bar{z}} \bar{\rho} - \theta^+ \theta^- \bar{\theta}^- \partial_z \bar{G} + \theta^- \theta^+ \bar{\theta}^- \bar{\theta}^+ \partial_z \partial_{\bar{z}} \bar{\rho}, \\
\overline{\mathcal{Q}}_+ \overline{\Psi} &= -2 \bar{\theta}^- \partial_{\bar{z}} \bar{\psi} + 2 \theta^- \bar{\theta}^- \partial_{\bar{z}} \bar{\rho} - 2 \theta^- \theta^+ \bar{\theta}^- \partial_z \partial_{\bar{z}} \bar{\psi}, \\ 
\overline{\mathcal{Q}}_- \overline{\Psi} &= \bar{\xi} + \theta^- \bar{G} + \theta^- \theta^+ \partial_z \bar{\xi} + \bar{\theta}^- \bar{\theta}^+ \partial_{\bar{z}} \bar{\xi} - \theta^- \bar{\theta}^- \bar{\theta}^+ \partial_{\bar{z}} \bar{G} + \theta^- \theta^+ \bar{\theta}^- \bar{\theta}^+ \partial_z \partial_{\bar{z}} \bar{\xi}.
\end{split}
\end{equation}

In order to obtain the usual supersymmetry transformations in the physics literature, we can define a total supersymmetry transformation operator:
\begin{equation}
\delta := \epsilon^+ \mathcal{Q}_+ + \epsilon^- \mathcal{Q}_- + \bar{\epsilon}^+ \overline{\mathcal{Q}}_+ + \bar{\epsilon}^- \overline{\mathcal{Q}}_-,
\end{equation}
where $\epsilon^{\pm}, \bar{\epsilon}^{\pm}$ are anti-commuting parameters. Then we can derive the usual supersymmetry transformations:
\begin{equation}
\begin{split}
\delta \phi &= \epsilon^+ \chi + \bar{\epsilon}^+ \lambda, \\
\delta \chi &= -2 \epsilon^- \partial_z \phi - \bar{\epsilon}^+ F, \\
\delta \lambda &= -2 \bar{\epsilon}^- \partial_{\bar{z}} \phi + \epsilon^+ F, \\
\delta F &= 2 \bar{\epsilon}^- \partial_{\bar{z}} \chi - 2 \epsilon^- \partial_z \lambda.
\end{split}
\end{equation}

\begin{equation}
\begin{split}
\delta \bar{\phi} &= \epsilon^- \bar{\chi} + \bar{\epsilon}^- \bar{\lambda}, \\
\delta \bar{\chi} &= -2 \epsilon^+ \partial_z \bar{\phi} - \bar{\epsilon}^- \bar{F}, \\
\delta \bar{\lambda} &= -2 \bar{\epsilon}^+ \partial_{\bar{z}} \bar{\phi} + \epsilon^- \bar{F}, \\
\delta \bar{F} &= 2 \bar{\epsilon}^+ \partial_{\bar{z}} \bar{\chi} - 2 \epsilon^+ \partial_z \bar{\lambda}.
\end{split}
\end{equation}

\begin{equation}
\begin{split}
\delta \psi &= \epsilon^+ \rho + \bar{\epsilon}^+ \xi, \\
\delta \rho &= -2 \epsilon^- \partial_z \psi - \bar{\epsilon}^+ G, \\
\delta \xi &= -2 \bar{\epsilon}^- \partial_{\bar{z}} \psi + \epsilon^+ G, \\
\delta G &= 2 \bar{\epsilon}^- \partial_{\bar{z}} \rho - 2 \epsilon^- \partial_z \xi.
\end{split}
\end{equation}

\begin{equation}
\begin{split}
\delta \bar{\psi} &= \epsilon^- \bar{\rho} + \bar{\epsilon}^- \bar{\xi}, \\
\delta \bar{\rho} &= -2 \epsilon^+ \partial_z \bar{\psi} - \bar{\epsilon}^- \bar{G}, \\
\delta \bar{\xi} &= -2 \bar{\epsilon}^+ \partial_{\bar{z}} \bar{\psi} + \epsilon^- \bar{G}, \\
\delta \bar{G} &= 2 \bar{\epsilon}^+ \partial_{\bar{z}} \bar{\rho} - 2 \epsilon^+ \partial_z \bar{\xi}.
\end{split}
\end{equation}

\end{document}